\newcommand{\be}{\begin{equation}}
\newcommand{\ee}{\end{equation}}
\newcommand{\bea}{\begin{eqnarray}}
\newcommand{\eea}{\end{eqnarray}}

\newcommand{\zs}{\sigma}
\newcommand{\zt}{\tau}

\newcommand{\deriv}[2]{{\frac{d#1}{d#2}}}

\documentclass[12pt]{article}

\begin{document}

\title{\bf About the  maximum entropy principle  in non equilibrium statistical mechanics}

\maketitle

\author{Gennaro Auletta\footnote{gennaro.auletta@gmail.com, University of Cassino},
Lamberto Rondoni\footnote{Corresponding author: lamberto.rondoni@polito.it: Dipartimento 
di Scienze Matematiche and Graphene@Polito Lab, Politecnico di Torino, Corso Duca degli Abruzzi 24, 
10129 Torino, Italy. INFN, Sezione di Torino, Via Pietro Giuria 1, 10125 Torino, Italy}
and Angelo Vulpiani\footnote{Angelo.Vulpiani@roma1.infn.it: Universit\`a di  
Roma ``Sapienza'', Dipartimento di Fisica, and 
CNR, Istituto Sistemi Complessi, P.le A. Moro 2, I-00185 Rome, Italy}}

\begin{abstract}
The maximum entropy principle (MEP) apparently allows us to derive,
or justify, fundamental results of equilibrium statistical mechanics. 
Because of this, a school of thought considers the MEP 
as a powerful and elegant way to make predictions in physics and 
other disciplines, which constitutes an alternative and more general
method than the traditional ones of statistical mechanics.
Actually, careful inspection shows that such a success is due to a 
series of fortunate facts that characterize the physics of equilibrium 
systems, but which are absent in situations not described by Hamiltonian dynamics,
or generically in nonequilibrium phenomena.
Here we discuss several important examples in non equilibrium statistical mechanics,
in which the MEP leads to incorrect predictions, proving that it does not have a  
predictive nature. We conclude that, in these paradigmatic examples, the ``traditional'' methods
based on a detailed analysis of the relevant dynamics cannot be avoided.
\end{abstract}

\vspace{2pc}
\noindent{\it Keywords}: Statistical inference, turbulence, dissipation, fluctuation relations


\section{Introduction}

Statistical mechanics has been constructed, in the second half of the 19-th century,
by Maxwell, Boltzmann and Gibbs on the basis of the assumed microscopic dynamics, and additional hypothesis
(as ergodicity).
On the other hand there is today a  radically anti-dynamical point of view, according to which
statistical
mechanics were nothing else but a form of statistical inference, rather than a theory
of objective physical reality. Here with {\em statistical inference} it is understood the process 
of deducing properties of an underlying probability by means of some general criterion. Under this 
light, probabilities measure
the degree of truth of a logical proposition about the state of the system, rather than describing
the state of a system as such.

In this context, Jaynes \cite{Jaynes_1,Jaynes_2,Jaynes_3} 
proposed the maximum entropy principle (MEP)
 as the general rule for finding the probability of a given 
event when only partial
information is available.
Let us briefly summarize the main points.
 If the mean values of $m$ independent
functions $f_i({\bf x})$, where $\bf x$ is a vector which describes the state of the system, are given:
\be
c_i = \langle f_i \rangle = 
\int f_i({\bf x})\rho({\bf x}) d {\bf x} \,\,\,\,, i=1, ... , m \,\,\, ,
\label{eqno1}
\ee
the MEP rule determines the probability density $\rho$ of the events
compatible with these mean values, by maxim{is}ing the ``entropy''
\be
H= - \int \rho({\bf x}) \ln \rho({\bf x}) d{\bf x} \,\, ,
\label{eqno2}
\ee
under the constraints $c_i= \langle f_i \rangle$. For independent functions $f_1, \ldots, f_m$ we mean that it is not possible to find $a_1, \ldots, a_m \neq 0$ such that 
\begin{equation}
\sum_{j=1}^ma_jf_j({\bf x}) = 0. 
\end{equation}
Using the maximization method of the 
Lagrange multipliers one easily obtains
\be
\rho({\bf x})= 
{1 \over Z} \exp \sum_{i=1}^m \lambda_i f_i({\bf x})
\label{eqno3}
\ee
where $\lambda_1, \lambda_2 ...\lambda_m$ depend on $c_1, c_2,
... , c_m$.  For instance, for systems with a fixed number of
particles subjected to the unique constraint that their mean energy is
fixed, the MEP leads to the canonical distribution in a very simple
fashion. 

As a technical but rather important detail{,} we note that the above result holds
only if ${\bf x}$ is the vector of the canonical coordinates
(i.e.\ positions and momenta of the particles). Analogously, for
systems of varying numbers of particles, the grand canonical
distribution is obtained {by }additionally constraining the mean
{number of} particles. 

Many find in these facts an unquestionable evidence for the validity of the MEP.
We do not share such an opinion:  the success of the MEP in deriving the correct 
probability distribution in equilibrium statistical mechanics is just a
matter of fortunate coincidence, related to the choice of canonical
coordinates. Thus, the
weakest technical aspect of the MEP approach is the dependence of the
results on the choice of the variables. 

There is another and more important objection to the MEP: our
ignorance cannot be credited to add knowledge about real phenomena. As a matter of fact, 
in spite of the optimistic claims of the MEP
enthusiasts, to the best of our knowledge MEP has only produced
different, although sometimes more elegant, derivations of 
results that were already previously known. 

For {simplicity's sake}, consider a scalar random variable $X$,
ranging over a continuum, whose probability distribution function is
$p_X$.  It is easy to real{is}e that the ``entropy'' 
\begin{equation}
H_X = - \int
p_X(x) \ln p_X(x) \, dx
\end{equation}
is not an intrinsic quantity of the phenomena
concerning $X$.  With a different parametrization, i.e.\ using the
coordinates $y=f(x)$ with an invertible function $f$, rather than $x$,
the entropy of the same phenomenon would now be given by
\be
H_Y = - \int p_Y(y) \ln p_Y(y) \, dy,
\label{eqno4}
\ee
with $p_Y(y)=p_X(f^{-1}(y))/|f'(x=f^{-1}(y)|$. Therefore, one has 
\be
H_Y=H_X + \int p_X(x) \ln|f'(x)| \, dx,
\label{eqno5}
\ee
so the MEP gives different solutions if different variables are
adopted to describe the very same phenomenon.

In order to avoid such an unpleasant
 dependence on the choice of variables, Jaynes
later proposed a more {sophisticated} version of the MEP, in terms of
the relative entropy:
\be 
{\tilde H}=-\int \rho({\bf x})
 \ln \left[ {\rho({\bf x}) \over q({\bf x})} \right] d{\bf x} \,\ ,
\label{eqno6}
\ee
where $q$ is a given probability density. Of course, ${\tilde H}$
 depends on $q$; but, at variance with the entropy, it does not depend
 on the chosen variables. On the other hand, one must decide how to
 select $q$, and this issue is equivalent to the problem of choosing
 the ``proper variables''.  Therefore, even this more elaborate method
 is non-predictive, and we see no reason to pursue {the MEP approach
   further} in the field of statistical mechanics.
For a detailed discusson on the MEP in equilibriun statistical
mechanics see \cite{CHIBBARO,CHIB2,Uffink_1995,Uffink2,kn:Shimony_1985,Friedman_Shimony_1971}.
\\
The aim of the present paper is to discuss some non standard topics 
of statistical physics, namely fluid mechanics and non equilibrium
problems, showing how the statistical features are determined by precise
dynamical behavior and cannot be predicted (even at a qualitative level) by inference ideas
as in the MEP.
\\
In Section 2 we discuss the statistical mechanics of fluids:
only in the inviscid case, which is for many aspects similar
to the Hamiltonian systems, MEP is able to give the correct result.
On the contrary, in the more interesting situation, of turbulent
flows MEP is not able to select the correct statistical features
which are selected only  by some physical aspects of the dynamics.
\\
Section 3 is devoted to the  non equilibrium statistical mechanics.
In some cases the MEP can predict the proper results but only
using  the relevant variables and constraints.
Such assumptions, in terms of inference, are not natural at all.
In particular the claimed success of the MEP for the fluctuation
relations is due to serious confusion between formulae that are only 
apparently similar while, in reality, they describe completely different 
physical situations. The last Section is devoted to our concluding remarks.

\section{Statistical fluid mechanics and maximum entropy principle}

In spite of the fact that, in general, a fluid (even in absence of viscosity)
does not obey Hamiltonian equations, it is easy to develop an equilibrium
statistical mechanical treatment for the Euler equation in perfect
analogy with the micro-canonical formalism used in standard
Hamiltonian systems.

Let us consider a perfect fluid, i.e. with zero viscosity,
and without external forcing, in a cube of edge $L$ with periodic boundary
conditions, so that the velocity field  can be expanded in Fourier
series as
\be
 u_j({\bf x},t)=L^{-3/2}\sum_{n_1,n_2,n_3} e^{i(k_1x+k_2y+k_3z)} v_j({\bf k},t)
\label{eqno7}
\ee
where
$$
{\bf k}={2\pi \over L}(n_1,n_2,n_3)
$$
with $n_j$ integer numbers. We introduce an ultraviolet truncation $v_j({\bf k})=0$ for
$|{\bf k}|>K_M$, being $K_M$ the maximum allowed wave vector. 

Because of the incompressibility  condition ($\nabla \cdot {\bf u}=0$)
and the fact that the velocity
field ${\bf u}({\bf x},t)$  
is real the variables $\{ v_j({\bf k},t) \}$ are not independent, {\em e.g.} one has 
\begin{eqnarray}
\sum_{j=1}^3 k_jv_j({\bf k}, t) = 0 & {\rm and} & v_j(-{\bf k},t) = \left[v_j({\bf k}, t)\right]^\star,
\end{eqnarray}
where $\star$ denotes complex conjugation. Therefore it is useful to  introduce a new set of variable 
$\{ X_n(t) \}$ replacing $v_j({\bf k},t)$ and obeying an
ordinary differential equation:
\be
{d X_n \over dt}=\sum_{m,\ell}M_{n,m,\ell}X_m X_{\ell} \,\, , \,\,
n=1,2,.... N \sim K_M^3\,\, .
\label{eqno8}
\ee
from the Euler's equation we have the following properties:
where $M_{n,m,\ell}=M_{n,\ell,m}$  and
$M_{n,m,\ell}+ M_{m,\ell,n}+ M_{\ell,n,m}=0$: for details see \cite{kn:Frisch_1995, Bohr}.
Because of the introduction of the ultraviolet truncation,
we have a finite system of equations, therefore one
avoids the infinite energy problems of the classical field
theory.

Since Eq. (\ref{eqno8}) conserves the volume in the phase space
(Liouville theorem)
\be
\sum_n{\partial \over \partial X_n} {d X_n \over dt}=0
\label{eqno9}
\ee
and in addition one has the (energy) conservation law
$$
{1 \over 2}\sum X_n^2=E \,\, ,
$$
it is straightforward, following the usual approach of 
equilibrium statistical mechanics, to derive the microcanonical
distribution:
\be
P_{mc}(\{ X_n \}) \propto \,\,\, \delta\Big( {1 \over 2}\sum X_n^2-E \Big) \,\, .
\label{eqno10}
\ee
In addition, the $N \to \infty$ limit yields the canonical distribution
\be
P_{c}(\{ X_n \}) \propto  \exp \left[- \Bigl( {\beta \over 2}\sum X_n^2 \Bigr)\right]
\label{eqno11}
\ee
and therefore
\be
<X_n^2>={2 E \over N}= {1 \over \beta}\,\,\,
\label{eqno12}
\ee
The previous procedure can be easily generalized to the
two-dimensional case in which there is a second conserved quantity,
the enstrophy (the quantity related to the kinetic energy in the flow model 
that corresponds to dissipation effects in the fluid):
\begin{equation}
\Omega={1 \over 2} \sum_n k_n^2 X_n^2 \,\, .
\end{equation}
Because of this, the microcanonical distribution should be defined on the
surface in which both energy and enstrophy are constant, and in
the large $N$ limit, we have the canonical distribution
\be
P_{c}(\{ X_n \}) \propto  \exp \left[- \Bigl(
 {\beta_1 \over 2}\sum X_n^2 +  {\beta_2 \over 2}\sum k_n^2 X_n^2   
\Bigr)\right]
\label{eqno13}
\ee
and therefore
\be
<X_n^2>={1 \over \beta_1+ \beta_2 k_n^2}\,\, .
\label{eqno14}
\ee
Detailed numerical simulations show that systems described 
by inviscid ordinary differential equations, such as Eq.(\ref{eqno8}), 
with quadratic invariants, for which the Liouville theorem holds, are ergodic 
and mixing if $N$ is large. Then arbitrary initial distributions of 
$\{ X_n \}$ evolve towards the Gaussian (\ref{eqno11}) or (\ref{eqno13}),
see \cite{Bohr}.

In the inviscid case, the Liouville theorem implies that 
the ``natural'' variables are $\{ X_n \}$, and the success of the MEP to derive
the canonical distribution is quite obvious.
The reason is the same as for the statistical mechanics
of Hamiltonian systems. We additionally remark that in usual statistical mechanics,
one only uses the energy conservation and the Liouville theorems,
while the full Hamiltonian structure plays no role.

Let us now discuss the more interesting case of real fluids,
where a viscosity and forcing are present.
Particularly interesting is the
fully developed turbulence,
where the Reynolds number $R_e=UL/\nu$ (being $U$ and $L$
the typical velocity and length respectively) is very high.
At variance with naive expectations on the statistical features of turbulence,
based on the incorrect assumption that $R_e \to \infty$ is equivalent to $\nu=0$,
the scenario is very different \cite{kn:Frisch_1995}. In $3D$, instead of equipartion, 
we have Kolmogorov's law i.e. $E(k_n) \sim k_n^2 <X_n^2> \sim k_n^{-5/3}$, therefore
\begin{equation}
<X_n^2>\sim k_n^{-11/3} \,\, .
\end{equation}
The previous law can be understood in terms of a cascade mechanism: see \cite{kn:Frisch_1995, Bohr}.

Let us note that, using MEP and imposing the (natural)  constraint
\begin{equation}
\sum_n<X_n^2>={\rm const},
\end{equation}
we obtain Eq.(\ref{eqno12}) {\em i.e.} the same result of the inviscid case.

Because of the presence of the viscosity and because of the inapplicability of the Liouville 
theorem, in fully developed turbulence there are no ``natural'' variables.

\subsection{Statistical features of turbulent models}

In order to understand of the difference between the cascade mechanism
and the ``equipartition'' scenario a numerical study is unavoidable. On 
the other hand, a numerical simulation of the Navier--Stokes equations in the 
limit $R_e >> 1$ is a prohibitive task. If the interest is only for a study of 
the scaling behavior, one can use simplified dynamical models, 
the so called shell models (SM), which, in spite of their apparent simplicity, 
reproduce many statistical features observed in experiments and in detailed 
numerical simulations \cite{Bohr,Ditle,Benzi}.

The basic idea of the SM is to implement a dynamical 
(energy or other quantities)  cascade model in terms of
a set of complex variables $u_n$, $n = 1, . . . , N$ representing the
velocity fluctuations in a shell of wave-numbers $k_n < |{\bf k}| < k_{n+1}$.
The wave-numbers are chosen geometrically spaced $k_n = k_0 2^n$ 
therefore  the number of variables needed to describe the inertial range
physics, is not too large.
In this way, the spatial and vectorial structure of the
original problem is completely disregarded. Then, some insights are
used to derive the equations ruling the set of variables $\{ u_n \}$.
A basic source of inspiration is the Navier-Stokes equation (NSE)
written in Fourier space, where the modes
interact in triads (see e.g. Eq. (\ref{eqno8})): only three modes are
involved at the same time. In this way, we simplify the complexity of
the equations by retaining the triad structure and eliminating
some interactions. Due to the hierarchical organization of the
characteristic times associated with the different scales,
we can assume that only close modes, {\em i.e.} variables
referring to close scales, can interact. The justification for this is
that distant modes (say $k_n$ and $k_m$ with $|m-n|\gg 1$) have so different
timescales that the resulting interaction would be very weak. This
assumption is known as the hypothesis of locality of the cascade, and
can be substantiated with refined analysis of the NSE.

According to the previous ideas we can introduce a set of
ordinary differential equations:
\begin{equation}
{du_n \over dt} =-\nu k_n^2 u_n + g^{(\alpha)}_n(u_n, u_{n \pm 1}, u_{n \pm 2}) + f_n,
\end{equation}
where  $f_n$  is the external forcing, the term  $-\nu k_n^2 u_n$
corresponds to the dissipation, while the term
$g^{(\alpha)}_n(...)$ includes the nonlinear terms,
and the parameter $\alpha$ determines the
conservation laws in the inviscid limit $\nu=0 \, , \, f_n=0$:
\begin{equation}
{du_n \over dt} =-\nu k_n^2 u_n +
ik_n\Bigl(a_nu^{*}_{n+1} u^{*}_{n+2}+
{b_n \over 2} u^{*}_{n-1} u^{*}_{n+1}+
{c_n \over 4} u^{*}_{n-1} u^{*}_{n-2}
\Bigr) +f_n \,\, ,
\end{equation}
with $n=1, ... ,N$, $b_1=b_N=c_1=c_2=a_N=a_{N-1}=0$. 

Given the conservation of energy $\sum_n|u_n|^2$ when
$\nu=f_n=0$, one has the constraint $a_n+b_{n+1}+c_{n+2}=0$, and the time scale 
can be fixed applying the condition $a_n=1$. This leaves one free 
parameter $\delta$:
\begin{equation}
a_n=1\,\, , \,\, b_n=-\delta\,\, , \,\, c_n=-(1-\delta)\,\, . 
\end{equation}
In the inviscid limit, the systems possesses a second conserved quantity:
\begin{equation}
\sum_nk_n^{\alpha} |u_n|^2 \,\, ,
\end{equation}
where $\alpha$ and $\delta$ are linked by the relation
$2^{\alpha}=1/(1-\delta)$ \cite{Bohr}.
The cases $\delta=1/2$ and $\delta=5/4$ correspond
the $3d$ and $2d$ turbulence respectively.

In spite of their (apparent) naive character, the shell models
are non trivial at all and maintain  all the difficulties of the NSE.
Remarkably for $\delta=1/2$ the shell model shows the same rich
statistical features observed in labs and direct numerical simulations 
of the NSE, e.g. the anomalous scaling of the structures functions
and the shapes of the probability distribution of many relevant quantities.
Let us note that the agreement holds also at the quantitative level.

The great advantage of shell models is that the number of shells $N$
necessary to mimic the cascade mechanism of fully developed turbulence
is relatively small, because of the geometrical progression in $k_ n$ we
roughly have $N \sim \ln (R_e)$.
 We have thus a chaotic dynamical system
with a reasonably small number of degrees of freedom where methods of
deterministic chaos can be used to link the statistical description to
the dynamical properties.

In the past years  shell models attracted the attention of many
scientists with different aims: the possibility to perform detailed
numerical computation on a model for the energy cascade to test ideas
or conjectures, e.g. in the context of predictability. Also, they have been used
to investigate analytic methods (to test some ideas for the closure problem),
developing rigorous results,
understanding the link between dynamical properties in phase space and
more standard quantities (in traditional turbulent literature) such as
structure functions and velocity  probability distribution.

Although only the cases  $\delta=1/2$ and $\delta=5/4$ correspond to
real physical situations ($3D$ and $2D$, respectively), it is interesting
to study the model also for other values of $\delta$: see \cite{Bohr}.
Let us discuss only the case $\delta >1$ corresponding to
real $\alpha$.

In order to determine the main statistical features,
we can follow two different statistical arguments:
\begin{itemize}
\item[a)] 
equipartition, i.e. $(k_n^{\alpha}+{\rm const}.) <|u_n|^2>={\rm const}$ which, for
for large $k_n$ implies
$<|u_n|^2> \sim k_n^{-2 \zeta}$ where $\zeta=\alpha/2$
\item[b)] cascade  \'a la Kolmogorov (see e.g. \cite{Bohr}), obtaining
$<|u_n|^2> \sim k_n^{-2 \zeta}$, where $\zeta=(\alpha+1)/3$
\end{itemize}
Actually, numerical simulations (see \cite{Bohr,Ditle}) show, apart small corrections due to intermittency, 
$\zeta=(\alpha+1)/3$ if  $\alpha <2$, and  $\zeta=\alpha/2$ for $\alpha >2$,
{\em i.e.}
\begin{equation}
\zeta=max\{{\alpha+1 \over3}, {\alpha  \over 2}\} \,\,.
\end{equation}
In short, we can say that for $\alpha <2$ 
the most important mechanism is the cascade (\`a la Kolmogorov),
while for  $\alpha >2$ the equipartition mechanism has the leading role for the scaling.
Let us present the physical argument \cite{Bohr}:
neglecting intermittency we assume  the simple scaling 
$u_n\sim k_n^{-h}$, therefore by dimensional arguments,
 the typical time at scale $k_n$ is $\tau_n \sim u_n k_n \sim k_n^{h-1}$.
Assuming a generalized $\alpha$-entrophy cascade one has a
constant rate for the generalized $\alpha$-enstrophy transfer,
i.e. $k_n^{\alpha} |u_n|^2/\tau_n \sim const.$ Therefore one
obtains $h=(1+\alpha)/3$ corresponding to  $\zeta=(\alpha+1)/3$.

In the previous argument one has
$$
\tau_n \sim k_n^{(\alpha-2)/3} \,\,,
$$
such a scaling is contradictory for $\alpha \ge 2$,
because it implies that the turn-over time $\tau_n$
does not decrease as $k_n^{-1}$ decreases.
Therefore we have an unrealistic result:
it is not possible to stop the cascade with a dissipative
mechanism whose characteristic time is $\tau_n^{(d)} \sim k_n^{-2}$,
i.e., at variance with the case $\alpha < 2$,
it is not  possible to find a $k_{diss}=k_{n^*}$ such that
$$
\tau_{n^*} \sim \tau_{n^*}^{(d)} \,\, .
$$
As consequence of the failure of the cascade mechanism, for
$\alpha \ge 2$, it is reasonable to expect an
equilibrium statitical scenario, with $\zeta=\alpha/2$.

The previous phenomenological argument is well confirmed
by numerical computations, see \cite{Bohr,Ditle}.

The  above results on the dependence on $\zeta$ from $\alpha$
 originate from genuine physical arguments
and it is hard to believe that they could be obtained by mere inference arguments.

\section{Non--equilibrium examples}

In this section we discuss the MEP approach in the case of non-equilibrium systems.  
Dewar \cite{Dewar,Dewar2} claimed that the MEP
can be used to obtain the probability distributions for general non--equilibrium systems.

For sake of simplicity, let us consider discrete times, 1,2,3,... and let
\begin{equation}
\mathcal{T}_N = \left\{ x_1, x_2, \dots x_N \right\}
\label{nn1}
\end{equation}
be a trajectory segment of length $N$ in the phase space or state space of a given system of interest. One would 
like to identify the steady state probability density $p({\cal T}_N)$ about this trajectory segment in the 
state space of trajectory segments. The method stemming from the MEP relies on the maximization of the corresponding
``entropy'':
\begin{equation}
H_N= -\int  p(\left\{ x_1, x_2, \dots x_N \right\}) \ln p(\left\{ x_1, x_2, \dots x_N \right\}) ~ d x_1 \dots d x_N
\label{nn2}
\end{equation}
under the constraints concerning $M$ observables:
\begin{equation}
c_{j,N}= < f_{j,N} > \,\,\, , \,\, j=1,2, ..., M 
\label{nn3}
\end{equation}
where $f_{j,N}=f_{j,N}(\mathcal{T}_N)$: the scalars $c_{j,N}, f_{j,N}$ play the same roles as $c_j,f_j$ did in Section 1, and the notation stresses the fact that they 
refer to trajectories of length $N$ in the original phase space or state space. 
The MEP immediately leads to
\begin{equation}
p({\cal T}_N)= e^{- \sum_{j=1}^M  \lambda_j f_j({\cal T}_N)}
\label{nn4}
\end{equation}
where the values $\{ \lambda_1, \lambda_2, .. , \lambda_M \}$ are determined by those of $\{ c_{1,N}, c_{2,N}, .., c_{M,N}\}$ 
in Eq. (\ref{nn3}). 

In this respect, the MEP method does not differentiate equilibrium from non--equilibrium steady states; once the trajectory
of interest is identified, everything proceeds in the same way for both situations. This is indeed in line with considering 
the maximum entropy as a inference technique, which is then expected to work regardless of the physics and of the specific 
properties of the objects under investigation.
The difficulties one meets in such a completely general approach to non--equilibrium systems are as above:
\begin{itemize}
\item[\bf a)] the identification of the stationary state, {\em i.e.} of the suitable variables for describing it;
\item[\bf b)] the identification of the observables, {\em i.e.} of the relevant functions $\{f_j\}$ in Eq. (\ref{nn3}).
\end{itemize}
Cover and Thomas \cite{ThomasCover} in Chapter 11 of their well known book,
 express these ideas quite plainly:

\noindent
{\em Implicit in the use of the maximum entropy methods in physics is a sort of AEP
(asymptotic equipartition property) that says that all the micro states are equally probable.}

\noindent
In Section 12.6 of \cite{ThomasCover}, the reader can find a clear
discussion  of Burg's maximum entropy theorem, 
which states that the stochastic process  $\{ x_1, x_2, .... , x_N \}$ satisfying 
\begin{equation}
<x_n x_{n+k}>=C_k\,\,\,\, , \,\, k=1,2, .., p,
\label{nn5}
\end{equation}
where the correlation function $C_k$ is known, 
and enjoying the maximum entropy rate
\begin{equation}
h= \lim_{N \to \infty} {1\over N} H_N \,\, ,
\end{equation}
is the Gaussian Markov process obeying
\begin{equation}
x_n=\sum_{k=1}^p A_k x_{n-k} + \sigma z_n,
\label{nn6}
\end{equation}
with $\{ z_n \}$ i.i.d.\ Gaussian variables of zero mean and unitary variance, and  
$\{ A_k \}$ and $\sigma$ chosen so that Eq.(\ref{nn5}) is satisfied. Then, one has
\begin{equation}
p(\{ x_1, x_2, .... , x_N \}) = {1 \over K} \times  \exp - {1 \over 2 \sigma^2}\sum_n\left(x_n- \sum_p A_p x_{n-p}\right)^2,
\label{nn7}
\end{equation}
where $K$ is the normalization constant.

Arguably, in the simplest case one assumes
\begin{equation}
\langle x_n x_{n+k} \rangle = a^k <x^2>\,\,\,\, , \,\, \mbox{ with } \,\, 0<a<1,
\label{nn8}
\end{equation}
and the process maximizing $h$ is a discrete time Langevin equation, of form:
\begin{equation}
x_n=a x_{n-1} + \sigma z_n
\label{nn9}
\end{equation}
with $\sigma^2=<x^2>(1-a^2)$. As well known, the corresponding probability density $p$ is then given by \cite{Bettolo}
\begin{equation}
p(\{ x_1, x_2, .... , x_N \})= {1 \over K} \exp \left[- {1 \over 2 \sigma^2}\sum_n\left(x_n- ax_{n-1}\right)^2 \right].
\label{nn10}
\end{equation}
Despite being technically simple, this result reveals the serious limitations in which the MEP inevitably incurs.

For instance, Dewar \cite{Dewar,Dewar2} using the MEP approach obtains
\begin{equation}
p(\{ x_1, x_2, .... , x_N \})= e^{- \sum_j  \sum_n \lambda_j g_j(x_n)}
\label{nn11}
\end{equation} 
but with values $\{ \lambda_1, \lambda_2, .. , \lambda_M \}$ derived from constraints that are 
sums of functions of the variable $x$ at a given time. Consequently, one cannot account for a sum of
functions of $x_n$ and $x_{n-1}$. One may insist, and define the state at time $n$ in terms of a two
components array such as  ${\bf y}_n=(x_n, x_{n-1})$. In that case, indeed, the MEP approach might
work, if proper constraints are imposed, and one could claim that this is done in analogy with the
celebrated Onsager-Machlup work \cite{OM53}. In this paper on fluctuations and irreversible processes,
the authors pose a difficult question:

\noindent
{\it how do you know you have taken enough variables, for it to be Markovian? }

\noindent
Similarly, Ma \cite{1985}, page 29, observes:

\noindent
{\it  the hidden worry of thermodynamics is: we do not know how many coordinates
or forces are necessary to completely specify an equilibrium state.}

But it is rather plain that the analogy does not stand,
because the state at time $n$ for the present system is merely given by $x_n$, and there is 
no {\em a priori} reason to adopt ${\bf y}_n=(x_n, x_{n-1})$ as a description of the state.
One could equally reasonably opt for any other pair of variables. In a sense, such choices would
be like deciding that the harmonic oscillator is described by $x, dx/dt$ and $d^2x/dt^2$, instead of 
$x, dx/dt$ only.

\subsection{Fluctuation relations}
Dewar \cite{Dewar,Dewar2} used the information theoretic approach to non--equilibrium statistical mechanics 
also in order to derive one of the most popular results of the past decades: the fluctuation relations (FR) which deals with the probabilities
of a trajectory and its (time) reversed one. 
This is a symmetry relation of the probability of second law ``violating'' phase-space paths \cite{ESR,Mejia,Bettolo}.
Dewar's derivation seems to imply that the FR is a generic property of the MEP's probability distributions, 
involving constraints
on anti-symmetric functions, independently of any physical interpretation that may be associated to the phenomenon 
and to
the constraints. Physically, it would then suffice to apply the MEP to the entropy production of those macroscopic fluxes 
that vary under the imposed constraints, and that would amount to selecting the most probable macroscopic flux configuration. 
In this case, one denotes by $f_k$ the thermodynamic fluxes contributing to the entropy production. 

Let $\mathcal{T}_N^{(+)}$  be a trajecotry of length $N$ and probability $p_{\mathcal{T}_N^{(+)}}$, along which there 
is a positive entropy production, expressed by $\zs = \sum_{k=1}^m \lambda_k f_k(\mathcal{T}_N^{(+)})$. Such a trajectory
can be paired with a trajectory $\mathcal{T}_N^{(-)}$ of probability $p_{\mathcal{T}_N^{(-)}}$, corresponding to the opposite 
entropy production, $-\zs$. 
Using MEP, Dewar then obtains the following relation --- cf.\  Eq.(12) in Ref.\cite{Dewar,Dewar2} ---
\be
\frac{p_{\mathcal{T}_N^{(+)}}}{p_{\mathcal{T}_N^{(-)}}} =
\exp \left\{ \sum_{k=1}^m \lambda_k f_k\left(\mathcal{T}_N^{(+)}\right)
\right\},
\label{nn12}
\ee
which is apparently very general, because rather than entropy production for a non--equilibrium thermodynamic systems, one could have 
considered any process with $n$ outcomes $\{1,...,n\}$, whose events can be grouped in pairs $(i+,i-)$, such that the $f_k$'s obey 
$f_k(i-)=-f_k(i+)$. The result would have been identical, proving the incredibly general applicability of the FR. 

Dewar then observes that: {\em a common explanation for these relationships lies in the hypothesis that 
the trajectories have a Gibbs-type probability distribution. Maximal Entropy provides the natural formalism in which 
Gibbs-type distributions emerge, whether or not they refer to physical systems. Thus the fluctuation theorem 
is not confined 
to physical systems alone but arises in a (potentially large) class of statistical inference problems 
involving constraint functions which are anti-symmetric.}

To connect Dewar's result with the FR of non--equilibrium statistical physics, one has to take a sample space whose $n$ 
elements are the possible trajectories of a non--equilibrium system. This implies that the state space (for a stochastic 
process) is finite, or that the phase space (for a deterministic system) admits a finite generating partition.

Granting all that, it nevertheless appears that Dewar's Eq. (\ref{nn12}) does not distinguish the numerous different 
situations that may arise, and always yields the same expression (\ref{nn12}) even in the cases in which it is 
incorrect. Therefore, this MEP approach has no predictive value. In particular, Eq. (\ref{nn12}) incurs in a systematic 
error which is far from harmless, because it puts on the same footings two physically completely different questions and the 
correspondingly different experiments: 

{\bf 1)} Measurements concerning the properties of 
non--equilibrium steady states; 

{\bf 2)} Measurements concerning the properties of equilibrium states. 

\noindent
This formally appears in the fact that, {\em even when it does hold}, a steady state FR does not look like 
Eq. (\ref{nn12}) in general, but it contains a correction term $c_N$ that must turn negligible when $N$ grows:
\be
\frac{p_{\mathcal{T}_N^{(+)}}}{p_{\mathcal{T}_N^{(-)}}} = 
\exp \left\{ \sum_{k=1}^m \lambda_k f_k(\mathcal{T}_N^{(+)}) + d_N \right\}.
\label{nn14}
\ee
In the standard cases in which the steady state FR holds, $d_N$ is of order $O(1)$ compared to the order $O(N)$ of 
the sum in Eq. (\ref{nn14}), hence it is indeed negligible for large $N$. 
However, $d_N$ is related to the decay of correlations of microscopic events 
in the steady state, and when the relevant correlations do not decay sufficiently fast, it may get large with $N$
producing expressions that do not resemble Eq. (\ref{nn12}) --- cf.\
\cite{Mejia,Bettolo,GC99,SRE,BGGZ} --- at variance with the MEP approach.\footnote{Note that the $N$ is not required 
to be large for the system to reach a steady state; $d_N$ is present within the steady state dynamics. Also,
one cannot consider the collection of infinitely long trajectories ($N=\infty$) because, apart from those corresponding 
to the average entropy production, their probabilities vanish and the ratio
on the left hand side of (\ref{nn14}) has no meaning.}

As a matter of fact, a relation like Eq. (\ref{nn12}) lacking the correction term $d_N$, can be verified, but as a transient 
relation. Transient relations are very interesting tools, 
used to obtain equilibrium properties of given collections of system, by doing non--equilibrium work on
them \cite{Bettolo}. In a sense, transient FR close the circle with the Fluctuation Dissipation Theorem which does the opposite,
obtaining non--equilibrium properties from equilibrium experiments.

One could then argue that a correction of order $O(1)$ on a term of order $O(N)$ should be neglected in general, and 
that Eq. (\ref{nn12}) could be accepted {\em in practice} in all circumstances. However, this is a gross error. Apart 
from the above observations, there are at least three further major differences in the physics described by steady state
and transient FR:
\begin{itemize}
\item there is no indication in the MEP procedure that $N$ should be large, therefore the
accuracy of the supposed approximation of the correct FR cannot be estimated;
\item the probability $p$ appearing in the transient FR is the equilibrium probability and not the steady state
probability, so that $p$ in Eq.(\ref{nn12}) intended as a transient relation and $p$ in Eq.(\ref{nn14}) are totally 
different objects;
\item transient FR describe the statistics of different experiments starting in the same equilibrium state but with 
different initial microscopic state, as in the case of protein stretching or of colloidal particles dragged in water. 
Consequently, transient FR do not need to describe any given single object, in general,
and they do not even need to tend to any steady state expression when $N$ grows. Differently, steady state FR describe 
the fluctuations of a 
single non--equilibrium system in its steady state \cite{ESR,Mejia,Bettolo}.
\end{itemize}
Furthermore, the most serious difficulty lies again with the choice of the functions $f_k$.
Analogously to the previous general discussion, also in the case of the fluctuation relations one should know beforehand 
the correct variables by which the state of a system must be described, as well as the relevant observables. 
Unfortunately, 
in many circumstances a proper set of variables does not even exist \cite{BG}, and one may pass from a situation 
in which the steady FR holds (correlations decay fast and $d_N$ turns negligible with increasing $N$) to one in which it does 
not ($d_N$ remains comparable to the other terms) by merely changing parameters which play no role in the MEP 
approach \cite{OSID,GRS}.

\subsection{A working example}

The above remarks for deterministic dynamics have stochastic counterparts. Therefore, let us conclude this section 
considering a simple model, in which similar difficulties are encountered, as first pointed out by Farago for
systems in unbounded potentials \cite{Farago}. 
In particular, let us consider an overdamped Langevin process, describing a Brownian particle, dragged in a liquid  by 
a moving harmonic potential with a constant velocity ${v}^*$, which is relevant {\em e.g.} for the optical trap 
experiment \cite{WSMSE}:
\begin{equation} \label {eq:brownian}
\deriv{x(t)}{t} = -\left[ x(t) - x^*(t) \right] + \zeta(t) ~.
\end{equation}
Here $x(t)$ is the  position of the particle at time $t$,  $x^*(t) = v^*t$ the
position of  the minimum of  the potential, $\zeta(t)$  is a white  noise term
representing the thermal bath, and $k_BT = 1$.  Then, the work done in a time $\zt$ is
\begin{equation} \label{eq:vzc-work}
W_\tau = - v^* \int_0^\tau [x(t) - x^*(t)] dt \ .
\end{equation}
In this context, Van Zon and Cohen \cite{vzc} considered the energy balance
\begin{equation} \label{eq:e-balance}
W_\tau = Q_\tau + \Delta U_\tau \ .
\end{equation}
where $Q_\tau$ is the dissipated heat and $\Delta U_\tau$ the potential energy of a
colloidal particle. They then observed that in a comoving frame $W_\tau$ is Gaussian 
with variance $2 \langle W_\zt \rangle$. As this property perists asymptotically in
$\tau$, they could conclude that the steady state FR holds for the total work.

Differently, the PDF of the potential energy is exponential at
equilibrium,  P($\Delta U)  \sim \exp(- {\rm const.}\Delta  U)$, and is expected to remain
exponential even away from equilibrium. Consequently, the small fluctuations of
heat are expected to coincide with those of the total work, because the
contribution of the potential energy is only $\mathcal{O}(1)$, while large heat
fluctuations are more likely to be generated by large fluctuations of the potential
energy, thus they are not distributed like work.

As a result, the expression in the large $\tau$ limit for the heat FR takes the 
standard form
\begin{equation} 
\label{eq:Q-fluc}
\frac{P_\tau(Q_\tau)}{P_\tau(-Q_\tau)} \approx e^{Q_\tau}
\end{equation}
only for $Q_\tau \in \left[ 0,\langle Q \rangle\right)$. For $Q_\tau \in \left[ \langle Q \rangle, 3 \langle Q \rangle \right)$,
there is a complicated nonlinear function of $Q_\tau$ and $W_\tau$ in the exponential, and for $Q_\tau > 3 \langle Q \rangle$ one 
eventually obtains
\begin{equation} 
\frac{P_\tau(Q_\tau)}{P_\tau(-Q_\tau)} \approx e^{2 \langle Q \rangle}
\end{equation}
where $\langle \cdot \rangle$  is the  steady state  average.

This result, due to the insufficiently rapid decay of the PDF of heat, as opposed 
to that of the PDF of work, means that two perfectly analogous quantities from the the MEP standpoint, the
work and the dissipated energy, are in fact described by two substantially different 
FR, at variance with the MEP predictions.

Furthermore, Baiesi et  al.\ \cite{Baiesi} generalized the result of \cite{vzc}, providing necessary
conditions for the potential $V$ and for its motion $x^*(t)$, which are required by
the total work to satisfy the steady state FR. In particular, numerical tests showed 
that the steady  state FR does not hold for the total work if $x^*$ moves at constant 
velocity and $V$ is not symmetric.  Similar observations are reported in 
\cite{dhar,germans}.

As one may obtain non-symmetric potentials by changing one parameter in the model 
of \cite{vzc}, without affecting the parity of the total work, here we have another 
example in which the MEP approach is bound to make incorrect predictions.

The fact is that there are infinitely many different forms for the FT, each of
which depends on specific details of the systems under consideration.
It would be quite a surprise that any method generically based on an equipartition property   
treats properly such a plethora of different situations, especially considering that a 
typical feature of non--equilibrium systems, even close to equilibrium, is the absence of equipartition
\cite{rare}. As shown in many papers, this fact affects in vastly different fashions different 
observables and, to date, we have no way to point out which observables are most affected, 
except by direct investigation.

\section{Concluding remarks}

Apparentlu, the MEP approach may look like an elegant and powerful way to make
statistical inference in non--equilibrium situations. However, detailed
analysis reveals a serious difficulty, which makes non-predictive the
MEP method: it allows the derivations of the correct
expressions only when they are already known.

On the other hand, in
the non--equilibrium cases, one cannot proceed without knowledge of the
dynamics, which tends to be highly complex. In other words, the constraints one should
impose, even in the cases in which this can be done, should be related
to the dynamics.

It is worth mentioning that the MEP approach is also adopted in situations more complex than those concerning physical problems such
as the ones discussed above. Biology, for instance, provides countless examples in which the details of the 
relevant dynamics are not understood, hence it is often claimed that uninformed inference, of the MEP kind, 
is necessary. However, the maximization of the entropy is not appropriate to describe a living organism, 
since living 
organisms are not in equilibrium with the environment and they are characterized by a degree of order higher than 
that of the environment \cite{Auletta,Auletta_2011a}. This kind of order persists in time thanks to the energy and 
matter exchange of one 
organism with its environment. In non--equilibrium situations, correlations prevent the system from reaching the
possible maximum entropy.  Studies such as \cite{Friston-etal2014,AURELL} show that
biomolecules are in general not in a state of maximal entropy
precisely due to correlations among different components, as it is
evident {\em e.g.} in the emergence of the so--called ternary or
quaternary structure of proteins from the initial codified segments of
amino acids, that ultimately allow biological functions.

One may thus argue that the MEP should be replaced by an analogous inference 
principle suitable to characterize this exchange. For instance, Prigogine's minimum entropy production principle
correctly describes the system-environment exchange for stationary states in the linear regime of
irreversible thermodynamics. The principle asserts that the steady--state configuration minimizes the entropy 
production \cite{kn:Prigogine_1947}. The principle also suggests that the evolution will promote organisms 
that minimize their entropy production in their own environment, rather than maximizing the entropy. These 
organisms should turn out to be precisely those that are able to exert more control on the environment. However, 
the minimum entropy production principle does not possess an absolute generality. It fails as a system is taken 
farther and farther away from equilibrium, {\em i.e.} when it is driven towards higher and higher
dissipations. Therefore it suffers from the same difficulties of the MEP and, analogously to the MEP, it cannot 
be used as a general inference principle.

%
%



\begin{thebibliography}{99}


\bibitem{Jaynes_1} Jaynes, Edwin T.\ {\it Information Theory and Statistical Mechanics},
Physical Review {\bf 106}: 620--630 (1957)

\bibitem{Jaynes_2} Jaynes, Edwin T.\ {\it Information Theory and Statistical Mechanics II}, 
Physical Review {\bf 108}: 171--90 (1957)

\bibitem{Jaynes_3} Jaynes, Edwin T.\ {\it Information Theory and Statistical Mechanics}, in K. Ford (Ed.), 
{\it Statistical Physics}, New York, Benjamin Inc.: 181--218 (1963)

\bibitem{CHIBBARO}
Chibbaro, S., Rondoni, L., and Vulpiani A. {\it Reductionism, Emergence and Levels of Reality: The Importance 
of Being Borderline}, {Heidelberg, Springer} (2014).

\bibitem{CHIB2}
Chibbaro, S., Rondoni, L., and Vulpiani A.
{\it On the Foundations of Statistical Mechanics: Ergodicity, Many Degrees of Freedom and Inference},
Comm. Theor. Phys. {\bf 62} 469 (2014)

\bibitem{Uffink_1995} {Uffink, J.} 
{\it Can the Maximum Entropy Principle be Explained as a Consistency Requirement?}
{Studies in History and Philosophy of Modern Physics} {\bf 26} {223} (1995).

\bibitem{Uffink2} {Uffink, J.} 
{\it The Constraint Rule of the Maximum Entropy Principle}, {Studies in History and Philosophy 
of Modern Physics} {\bf 27}, {47} (1996)

\bibitem{kn:Shimony_1985}
Shimony, Abner,
{\it The Status of the Principle of Maximum Entropy},
{\it Synthese} {\bf 63} 35. {\it Search for a Naturalistic Point of View}, Cambridge, University Press. 

\bibitem{Friedman_Shimony_1971}
{Friedman, K. and Shimny, A.}
{\it Jaynes's Maximum Entropy Prescription and Probability Theory},
{J. Stat. Phys.}{\bf 3} {381} (1971)

\bibitem{kn:Frisch_1995}
Frisch, U., {\it Turbulence: The Legacy of A. N. Kolmogorov}, Cambridge University Press (1995). 

\bibitem{Bohr} T. Bohr, M.H. Jensen, G. Paladin and A. Vulpiani,
{\it Dynamical Systems Approach to Turbulence}, Cambridge University Press (1998).

\bibitem{Ditle} P.D. Ditlevsen and I.A. Mogensen
{\it Cascades and statistical equilibrium in shell models of turbulence} Phys. Rev. E {\bf 53}, 4785 (1996) 

\bibitem{Benzi} M.H. Jensen, G. Paladin and A.Vulpiani
"Intermittency in a cascade molel for 3-dimensional turbulence"
Phys. Rev. A {\bf 43}, 798 (1991)

\bibitem{Dewar} {Dewar, R.} 
{\it Information Theory Explanation of the Fluctuation Theorem, Maximum Entropy Production and Self--Organized Criticality in 
Non--Equilibrium Stationary States}, {J. Phys.} {\bf A36} {631} (2003).

\bibitem{Dewar2} Dewar, R.
{\it Maximum Entropy Production and the Fluctuation Theorem},
{J. Phys.} {\bf A38} {L371} (2005)

\bibitem{ThomasCover} {Cover, T.M. and Thomas, J.A.}
{\it Elements of Information Theory}, {Wiley, New York (1991)}

\bibitem{OM53}
{Onsager, L. and Machlup, S.}
{\it Fluctuations and Irreversible Processes},
{Physical Review} {\bf 91} 1505 (1953)

\bibitem{1985} 
{Ma, Shang--Keng},
{\it Statistical Mechanics},
{Singapore, World Scientific} (1985)

\bibitem{ESR} D.J.\ Evans, D.J\. Searles, L.\ Rondoni, {\it Application of the Gallavotti-Cohen fluctuation 
relation to thermostated steady states near equilibrium}, Phys.\ Rev.\ E {\bf 71} 056120 (2005)

\bibitem{Mejia} L.\ Rondoni, C.\ Mej\`a-Monasterio, {\it Fluctuations in nonequilibrium statistical 
mechanics: models, mathematical theory, physical mechanisms}, Nonlinearity {\bf 20} R1 (2007)

\bibitem{Bettolo} 
U. Marini Bettolo Marconi, A. Puglisi, L. Rondoni and A. Vulpiani,
{\it Fluctuation-Dissipation: Response Theory in Statistical Physics}, Phys. Rep. {\bf 461}, 111 (2008)

\bibitem{GC99}  E.G.D.\ Cohen, G.\ Gallavotti, {\it Note on two theorems in nonequilibrium statistical 
mechanics}, J.\ Stat.\ Phys.\ {\bf 96}, 1343 (1999)

\bibitem{SRE} D.J.\ Searles, L.\ Rondoni, D.J.\ Evans, {\it The Steady State Fluctuation Relation for 
the Dissipation Function}, J Stat Phys {\bf 128} 1337 (2007)

\bibitem{BGGZ} F.\ Bonetto, G.\ Gallavotti, A.\ Giuliani, F.\ Zamponi, {\it Chaotic hypothesis, 
fluctuation theorem and singularities}, J.\ Stat.\ Phys.\ {\bf 123} 39 (2006)

\bibitem{BG} F.\ Bonetto, G.\ Gallavotti, {\it Reversibility, coarse graining and the chaoticity principle}, 
Commun.\ Math.\ Phys.\ {\bf 189} 263 (1997)

\bibitem{OSID} L.\ Rondoni, G.P.\ Morriss, {\it Large fluctuations and axiom-Cstructures in deterministically 
thermostatted systems}, Open Syst.\ Information Dynam.\ {\bf 10} 105 (2003)

\bibitem{GRS} G.\ Gallavotti, L-\ Rondoni, E.\ Segre, {\it Lyapunov spectra and nonequilibrium ensembles
equivalence in 2D fluid mechanics}, Physica D {\bf 187} 338  (2004)

\bibitem{Farago} J.\ Farago, {\it Injected power fluctuations in Langevin equation} 
J.\ Stat.\ Phys.\ {\bf 107} 781 (2002)

\bibitem{WSMSE} G.M.\ Wang, E.M.\ Sevick, E.\ Mittag, D.J.\ Searles, D.J.\ Evans, {\it Experimental demonstration of violations
of the second law of thermodynamics for small systems and short time scales}, Phys.\ Rev.\ Lett.\ {\bf 89} 050601 (2002)

\bibitem{vzc} R.\ van Zon, E.G.D.\ Cohen, {\it Stationary and transient work-fluctuation theorems for a dragged Brownian
particle}, Phys.\ Rev.\ E {\bf 67} 046102 (2003)

\bibitem{Baiesi} M.\ Baiesi, T.\ Jacobs, C.\ Maes, N.S.\ Skantzos, {\it Fluctuation symmetries for work and heat}, Phys.\ Rev.\ E
{\bf 74} 021111 (2006)

\bibitem{dhar} T.\ Mai, A.\ Dhar, {\it Nonequilibrium work fluctuations for oscillators in non-Markovian baths} Phys.\ Rev.\
E {\bf 75} 061101 (2007)

\bibitem{germans} R.J.\ Harris, R\`akos, M.\ Sch\"utz, {\it Breakdown of Gallavotti-Cohen symmetry for stochastic dynamics},
Europhys.\ Lett.\ {\bf 75} 227 (2006)

\bibitem{rare} L.\ Conti, P.\ De Gregorio, G.\ Karapetyan, C.\ Lazzaro, M. Pegoraro, M-\ Bonaldi, L.\ Rondoni 
{\it Effects of breaking vibrational energy equipartition on measurements of temperature in macroscopic 
oscillators subject to heat flux}, 
JSTAT P12003 (2013)

\bibitem{Auletta} 
{Auletta, G.}
{\it A Paradigm Shift in Biology?}
{Information} {\bf 1} {28} (2010)

\bibitem{Auletta_2011a}
{Auletta, G.}
{\it Cognitive Biology: Dealing with Information from Bacteria to Minds},
{Oxford, University Press}

\bibitem{kn:Prigogine_1947}
Prigogine, I.,
{\it Etude th\'{e}rmodynamique des ph\'{e}nom\`{e}nes irr\'{e}versibles}, 
Li\`{e}ge, Desoer (1947)

\bibitem{Friston-etal2014} 
{Friston, K. J., Sengupta, B., and Auletta, G.}
{\it Cognitive Dynamics: From Attractors to Active Inference},
{Proceedings of the IEEE} {\bf 102} 427 (2014)

\bibitem{AURELL} Christoph Feinauer ,Marcin J. Skwark , Andrea Pagnani, Erik Aurell
{\it Improving Contact Prediction along Three Dimensions}, PLOS Comp Biol October 9, 2014


\end{thebibliography}
\end{document}